\documentclass[12pt]{article}
\usepackage{epsf,epsfig}
\usepackage{cite}
\usepackage{amssymb}
\usepackage[dvips]{color}
\usepackage{amsmath}
\usepackage{amsfonts}
\usepackage{latexsym}
\usepackage{slashed}
\usepackage{mathrsfs}

\makeatletter
\renewcommand\section{\@startsection {section}{1}{\z@}%
                                   {-3.5ex \@plus -1ex \@minus -.2ex}
                                   {2.3ex \@plus.2ex}%
                                   {\normalfont\large\bfseries}}
\renewcommand\subsection{\@startsection{subsection}{2}{\z@}%
                                     {-3.25ex\@plus -1ex \@minus -.2ex}%
                                     {1.5ex \@plus .2ex}%
                                     {\normalfont\bfseries}}
\makeatother

\let\non\nonumber

\renewcommand{\theequation}{\arabic{section}.\arabic{equation}}

\newcommand{\be}{\begin{equation}}
\newcommand{\ee}{\end{equation}}
\newcommand{\bea}{\begin{eqnarray}}
\newcommand{\eea}{\end{eqnarray}}
\newcommand{\bean}{\begin{eqnarray*}}
\newcommand{\eean}{\end{eqnarray*}}

\newcommand{\bb}{\mathbb}

\newcommand{\lp}{\left(}
\newcommand{\rp}{\right)}
\newcommand{\lb}{\left[}

\newcommand{\C}[1]{$(\ref{#1})$}

\newcommand{\pp}[1]{\frac{\p}{\p #1}}

\newcommand{\half}{\frac{1}{2}}

\newcommand{\rt}{{\sqrt 2}}

\renewcommand{\a}{\alpha}

\renewcommand{\d}{\delta}
\newcommand{\g}{\gamma}
\newcommand{\gt}{{\tilde{\gamma}}}
\newcommand{\gtb}{{\bar{\tilde{\gamma}}}}
\newcommand{\gbar}{{\bar\gamma}}

\newcommand{\s}{\sigma}
\renewcommand{\lb}{\bar{\lambda}}
\newcommand{\w}{\omega}

\newcommand{\e}{\epsilon}

\renewcommand{\t}{\theta}
\renewcommand{\tt}{{\tilde{\theta}}}
\newcommand{\tb}{\bar{\theta}}
\newcommand{\tp}{{\theta^+}}
\newcommand{\tpb}{{\bar{\theta}^+}}
\newcommand{\tm}{{\theta^-}}
\newcommand{\tmb}{{\bar{\theta}^-}}
\newcommand{\psibar}{{\bar\psi}}
\newcommand{\psit}{{\tilde{\psi}}}
\newcommand{\psitb}{{\bar{\tilde{\psi}}}}
\newcommand{\chibar}{{\bar\chi}}
\newcommand{\phib}{\bar{\phi}}

\newcommand{\vp}{\varphi}

\newcommand{\G}{\Gamma}

\newcommand{\T}{\Theta}
\newcommand{\U}{\Upsilon}
\newcommand{\LL}{\Lambda}
\newcommand{\LB}{\overline{\Lambda}}

\renewcommand{\S}{\Sigma}

\newcommand{\F}{{\cal F}}

\newcommand{\N}{{\cal N}}
\newcommand{\V}{{\cal V}}

\renewcommand{\P}{{\cal P}}
\newcommand{\J}{{\cal J}}

\newcommand{\IC}{{\mathbb C}}

\newcommand{\IP}{{\mathbb P}}

\newcommand{\IR}{{\mathbb R}}
\newcommand{\IZ}{{\mathbb Z}}

\def\XT{{\widetilde X}}

\newcommand{\tr}{\textrm{Tr}}
\def\I1{\relax{\rm 1\kern-.38em 1}}

\def\ie{{\it i.e.}}
\def\eg{{\eg e.g.}}
\newcommand{\hc}{\mathrm{h.c.}}

\def\p{\partial}
\def\pb{{\overline \p}}

\newcommand{\D}{{\rm D}}
\newcommand{\DB}{\overline{\rm D}}
\newcommand{\CD}{{\cal D}}
\newcommand{\CDB}{\overline{\cal D}}

\newcommand{\zb}{\bar{z}}

\def\NLSM{nonlinear sigma model}

\def\susy{supersymmetry}

\def\CY{Calabi-Yau}
\def\FY{Fu-Yau}
\def\Ka{K\"{a}hler}
\def\nK{non-K\"{a}hler}
\def\TT{$(2,2)$}
\def\ZT{$(0,2)$}

\newcommand{\DI}{\Delta^{-1}}
\newcommand{\QI}{(Q^{-1})}

\begin{document}

\begin{titlepage}
\begin{flushright}
\today\ \\
HUTP-06/A0045 \\
MIT-CTP 3785
\end{flushright}
\vskip 1in
\begin{center}
{\Large Linear Models for Flux Vacua}

\vskip 0.5in Allan Adams$^{1,2}$, Morten Ernebjerg$^{2}$ and Joshua M. Lapan$^{2}$
\vskip 0.4in {\it $^{1}$ Center for Theoretical Physics \\ Massachusetts Institute of Technology \\ Cambridge, MA  02139 USA}
\vskip 0.2in {\it $^{2}$ Jefferson Physical Laboratory \\ Harvard University \\ Cambridge, MA  02138 USA}
\end{center}

\vskip 0.8in

\begin{abstract}
\noindent
We construct worldsheet descriptions of heterotic flux vacua as the IR limits of  $\N$=2 gauge theories.  
Spacetime torsion is incorporated via a 2d Green-Schwarz mechanism in which a doublet of axions cancels a one-loop gauge anomaly.
Manifest $(0,2)$ supersymmetry and the compactness of the gauge theory instanton moduli space suggest that these models, which include Fu-Yau models, are stable against worldsheet instantons, implying that they, like Calabi-Yaus, may be smoothly extended to solutions of the exact beta functions.  
Since Fu-Yau compactifications are dual to KST-type flux compactifications, this provides a microscopic description of these IIB RR-flux vacua.

\end{abstract}

\end{titlepage}

\section{Introduction}

It is a beautiful and frustrating fact of life that Calabi-Yaus have interesting moduli spaces.  On the one hand, the topology and geometry of their moduli spaces govern the low-energy physics of string theory compactified on a Calabi-Yau, so understanding their structure teaches us about four-dimensional stringy physics.  On the other, the resulting massless scalars are a phenomenological disaster.

Dodging this bullet 
has proven surprisingly difficult.  At the level of type II supergravity, beautiful work of KKLT and others\footnote{See in particular \cite{Michelson:1996pn,Gukov:1999ya,Kachru:2002he,Kachru:2003aw} for foundational work, and \cite{Douglas:2006es} for a complete review and further references.} demonstrates that a judicious choice of fluxes and branes wrapping suitable cycles in a fiducial Calabi-Yau can generate a scalar potential which fixes all moduli of the underlying CY.  However, since these type II flux vacua necessarily involve RR fluxes and other effects which are not amenable to worldsheet analysis, it is difficult to construct a microscopic description for them, and a sufficiently hard-nosed physicist could rationally wonder whether these vacua, in fact, exist.  

Duality provides a powerful hint.  For a large class of flux vacua, such as the KST models of \cite{Kachru:2002he}, there exists \cite{Becker:2002sx} a duality frame involving a heterotic compactification on a \nK\ manifold of $SU$(3)-structure with non-trivial gauge and NS-NS 3-form flux, $H$$\neq$0,
all of which is in principle amenable to worldsheet analysis.  A microscopic description of heterotic flux vacua would thus provide a microscopic description of the dual KST vacua.

Of course, there are excellent reasons that most work has focused on \Ka\ compactifications, which necessarily have $H=0$.  In particular,  only for \Ka\ manifolds does Yau's Theorem ensure the existence of solutions to the tree-level supergravity equations; the beautiful results of Gross \& Witten \cite{Gross:1986iv} and Nemachamsky \& Sen \cite{Nemeschansky:1986yx} then ensure that these classical solutions extend smoothly to solutions of the exact string-corrected equations.    When $H\neq 0$, the story is much more complicated, due in part to the absence of effective computational tools analogous to Hodge theory or special geometry for \nK\ manifolds, and in part to the tremendous analytic complexity of the Bianchi identity,
\be\label{eq:HBI}
dH = \a' \lp \mathrm{tr} R\wedge R - \tr F\wedge F \rp . 
\ee
Indeed, twenty years passed between Strominger's geometric statement of the supersymmetry constraints \cite{Strominger:1986uh} and the proof by Fu and Yau of the existence of a class of solutions to these leading-order equations \cite{Fu:2006vj}.  Moreover, since the Bianchi identity scales inhomogeneously with the global conformal mode, any solution has total volume-modulus fixed near the string scale, so such compactifications can {\it not} be described by  conventional, weakly-coupled NLSMs.   Whether these \FY\ solutions, like \CY s, can be smoothly extended to solutions of the exact string equations has thus remained very much unclear.

The purpose of this paper is to develop tools with which to study heterotic compactifications with non-vanishing $H$, \ie\ holomorphic vector bundles over \nK\ manifolds with intrinsic torsion satisfying \C{eq:HBI}.  Motivated by Fu and Yau, we focus on torus bundles
over \Ka\ bases, $T^{m}\to X \to S$, with gauge bundle $\V_{X}$ and NS-NS flux $H$ turned on over the total space $X$.  When $m=2$ and $S=K3$, this is precisely the Fu-Yau compactification.\footnote{While we refer to these geometries as \emph{Fu-Yau} geometries, it should be emphasized that Strominger's elaboration of the precise equations to be solved was crucial to the eventual construction of solutions by Fu and Yau, as well as to earlier studies of the underlying manifolds by Goldstein and Prokushkin \cite{Goldstein:2002pg}.}

Our strategy closely parallels the familiar gauged linear sigma model (GLSM) approach to \CY\ compactifications\cite{Witten:1993yc}: we build a massive 2d gauge theory which flows in the IR to an interacting CFT with all the properties that we expect of a \FY\ compactification.  
In the \CY\ case, the GLSM flows to a NLSM whose large-radius limit is the chosen \CY.  This is not possible in the \FY\ case as no large-radius limit exists; however, the classical moduli space of the one-loop effective potential of our GLSM will precisely reproduce the \FY\ geometry.  We thus take the CFT to which our torsion linear sigma model (TLSM) flows to provide a microscopic definition of the \FY\ compactification.

A central ingredient in these models is a two-dimensional implementation of the Green-Schwarz mechanism.  The $ch_{2}(T_{S})-ch_{2}(\V_{S})$ anomaly of a $(0,2)$ \NLSM\  on $\V_{S} \to S$ is contained\footnote{In fact, this is a somewhat subtle story, as we shall elaborate below.} in the gauge anomaly of $(0,2)$ GLSMs for $S$.  In compactifications with intrinsic torsion, this sum does not vanish even in cohomology.   To restore gauge invariance, we introduce a novel $(0,2)$ multiplet containing a doublet of axions whose gauge variation precisely cancels the gauge anomaly.  The one-loop geometry of the resulting model is easily seen to be a $T^{2}$ fibration $X$ over the Calabi-Yau $S$ -- a Fu-Yau geometry -- with the anomaly cancellation conditions of the TLSM reproducing the conditions for the existence of a solution to the Bianchi identity.

Crucial to our construction is a manifest $(0,2)$ supersymmetry with non-anomalous $R$-current and a non-anomalous left-moving $U(1)$.  These ensure the perturbative non-renormalization of the superpotential and are necessary for the existence of a chiral GSO projection.  The worry, as usual in a $(0,2)$ theory, is that worldsheet instantons may generate a non-perturbative superpotential \cite{Dine:1986zy, Dine:1987bq}.  The power of a gauged linear description is that the moduli space of worldsheet instantons is embedded within the moduli space of gauge theory instantons, which is manifestly compact; without a direction along which to get an IR divergence, it is thus impossible to generate the poles required for the generation of a spacetime superpotential\cite{Silverstein:1995re,Beasley:2003fx}.  Such arguments have been used to rigorously forbid the existence of non-perturbative superpotentials for $(0,2)$ gauged linear sigma models of Calabi-Yau geometries; while some technical details differ so that we cannot present a direct proof, these results appear to extend unproblematically to our torsion linear models.

Along the way we will construct a number of $(2,2)$ TLSMs for generalized \Ka\ geometries, including non-compact models built out of  chiral and twisted chiral multiplets and more intricate models in which we gauge chiral currents built out of semi-chiral multiplets.  
While not our main interest in this paper, these models provide useful guidance in our construction of $(0,2)$ models with torsion and are worth studying on their own merits.

This paper is a brief introduction to the structure of torsion linear sigma models, focusing on a few basic results and proofs-of-principle.  A self-contained follow-up including examples and details omitted below is in preparation.

\section{Torsion in $(2,2)$ GLSMs}
\setcounter{equation}{0}

While a number of $(2,2)$ gauged linear sigma models with non-trivial NS-NS flux have been studied in the literature -- most notably the (4,4) $H$-monopole GLSM \cite{Tong:2002rq} -- the structure of general models has received relatively little attention.  In this section, we will review the incorporation of NS-NS flux into $(2,2)$ models, emphasizing features which will generalize to the more complicated $(0,2)$ examples studied below.  A more complete discussion of the rich structure of general $(2,2)$ torsion will be addressed elsewhere.

Let us start with a standard $(2,2)$ GLSM for some toric variety $V$ built out of chiral and vector supermultiplets.  The IR geometry of such models is necessarily \Ka.  What we seek is a way to introduce non-trivial $H=dB \neq 0$ into a standard $(2,2)$ GLSM.  Since $H$ is an obstruction to K\"{a}hlerity, we are also looking for a construction of \nK\ geometries via $(2,2)$ GLSMs.  It has long been known that sigma models built entirely out of chiral multiplets are necessarily \Ka\ \cite{Gates:1984nk}, so we would seem to need to introduce non-chiral multiplets.  However, since a $(2,2)$ gauge field minimally coupled to chiral multiplets cannot be minimally coupled to twisted chirals while preserving $(2,2)$, there would seem to be a no-go argument forbidding minimally-coupled GLSMs for \nK\ geometries with non-vanishing $H$.

As is often the case, this no-go statement tells us exactly where to go. Recall that $B$ appears in the GLSM through the imaginary parts of the complexified FI parameters $t^{a}=r^{a}+i\t^{a}$ appearing in the twisted chiral superpotential,
\[
- \frac{1}{2\sqrt{2}}\int d^{2}\tt ~ t^{a} \S_{a} +\textrm{h.c.}   ~ = ~ -r^{a}D_{a} + 2\t^{a} v_{+-a}.
\]
More precisely, $t^{a}$ are the restriction of the complexified Kahler class $\J=J+iB$ to the hyperplane classes $H_{a} \in H^{2}(V)$ corresponding to the gauge fields $\S_{a}$, \ie\ $B=\theta^{a} H_{a}$. To get $H\neq 0$ we must promote some of the $\t^{a}$, say $m$ of them, to dynamical fields. Note that this {\em adds} dimensions to the geometry, so we are no longer working with a sigma model on $V$, but with a geometry with local product structure $V\times(S^{1})^{m}$.  

For the moment, consider promoting a single FI parameter to a dynamical field.  Since the FI parameter appears in the twisted chiral superpotential, $(2,2)$ \susy\ requires that it be promoted to a twisted chiral multiplet $Y$ with action 
\be
-  \frac{N^a}{2\sqrt{2}}\int d^{2}\tt ~   Y \S_a  \, + \, \textrm{h.c.}    -\frac{1}{8}\int d^{4}\t ~   k^{2}(Y+\overline{Y})^{2}  
=  -k^{2}[(\p r)^{2} +(\p \t)^{2}] ~- N^a\left[r D_a   - 2\t ~\! v_{+-a}\right] + ... 
\ee
where $k\in\IR$ and $y=r+i\t\in\IC^{*}$ is the scalar component of $Y$.
The geometry is thus a complex manifold with local product structure, $W\sim V \times \IC^{*}$, and NS-NS potential $B=\t N^a H_a$ on the total space $W$ that is no longer closed,
\[
H=d\t \wedge N^a H_a \neq 0.
\] 
The resulting IR geometry is \nK,  evading the no-go statement above by coupling the gauge supermultiplet {minimally} to chirals and {\it axially} to twisted chirals.  Note that the resultant $H$-flux has two legs along $V$ and one along the $S^{1}$ coordinatized by $\t$.  Note, too, that this is precisely the form of the relevant couplings in the (4,4) $H$-monopole GLSM.

In some sense, what we have done by promoting the FI parameter/K\"{a}hler modulus $t$ to a dynamical field $Y$ is to take the variety $V$ and construct a new variety $W$ as a fibration of $V$ over a complex line in the K\"{a}hler moduli space of $V$.  This should give us pause; the moduli space includes points where the original variety $V$ goes singular, so this fibration is degenerate.  How do we know that the total space of the fibration is, in fact, smooth?

Consider, for example, the resolved conifold $F=xy-wz-r=0$ in $\IC^{4}$, and let $W$ be the fibration of the conifold over the complex line $r$.  The point $r=0$ is a very singular point -- even the CFT is singular -- and it is natural to wonder if $W$ is singular at $r=0$.  In fact, it is straightforward to see that $W$ is completely well behaved at $r=0$.  Like $V$, $W$ is the vanishing locus of $F$, now viewed as a function on $\IC^{5}$.  However, since $\p_{r}F=-1$, $F$ is strictly transverse, so the hypersurface $W=F^{-1}(0)$ is everywhere smooth.  By virtue of the linear nature of the axial coupling, a similar result can be argued to obtain for all $(2,2)$ models in which the FI parameter is promoted to a dynamical field.

T-dualizing the dynamical FI parameter is revealing.  Consider a GLSM with gauge group $U(1)^{s}$, $(N+s)$ chirals $\Phi_{I}$, and $m$ axially coupled twisted chirals, $Y_{l}$, with Lagrangian,
\[
{\cal L} = \int d^{4}\t ~\left[    -\frac{1}{4e_{a}^{2}}\overline{\S}_{a}\S_{a}  ~+ \frac{1}{4}\overline{\Phi}_{I}e^{2Q_{I}^{a}V_{a}}\Phi_{I}  ~-\frac{1}{8} k_{l}^{2}(\overline{Y}_{l}+Y_{l})^{2}\right]  ~ - \frac{1}{2\rt} \int d^{2}\tt ~  M^{a}_l Y_{l}\S_{a}  ~+\hc .
\]
Dualizing all the twisted chirals $Y_{l}$ into chiral multiplets $\P_{l}$ results in a simple model, 
\[
\tilde{{\cal L}} = \int d^{4}\t ~    \left[    -\frac{1}{4e_{a}^{2}}\overline{\S}_{a}\S_{a}  ~+ \frac{1}{4}\overline{\Phi}_{I}e^{2Q_{I}^{a}V_{a}}\Phi_{I}  ~+ \frac{1}{8k_l^{2}}(\overline{\P}_{l} +\P_{l}+2M^{a}_l V_{a})^{2}\right] .
\]
All matter fields are now chiral, so the classical moduli space is automatically \Ka.  But with which \Ka\ metric, on which space?  We can clearly eat the imaginary component of all $m$ fields $\P_{l}$ to make $m$ of the gauge fields massive (provided $s\geq m$), and use their real components to solve $m$ of the D-terms.  However, integrating out the massive vectors and scalars deforms the \Ka\ potential for the $(N+s)$ $\Phi_{I}$s.  The surviving $(s-m)$ gauge fields then effect a \Ka\ quotient of $\IC^{N+s}$, but now starting with a deformed \Ka\ structure.  The IR geometry is thus an $N+m$ dimensional variety whose topology is controlled by the charges $Q^{a}_{I}$ of the $\Phi_{I}$ under the surviving gauge fields but with deformed \Ka\ structure\cite{Hori:2002cd}.  This can be used to construct GLSMs for, say, squashed spheres.  T-dualizing with this squashed metric then gives non-trivial $B$, which was what we found above.


It is fun to note in passing that we could just as well have dualized the {\it chiral} multiplets in our torsion model to get a theory of only {\em twisted chiral} multiplets, all axially coupled to an otherwise free gauge multiplet.  As emphasized by Morrison \& Plesser \cite{Morrison:1995yh} and by Hori \& Vafa \cite{Hori:2000kt}, the resulting theory has a non-perturbative superpotential of the form $W=e^{-Z_{I}}$, where $Z_{I}$  are the twisted chirals dual to the original $\Phi_{I}$.  The resulting theories end up looking like complicated generalizations of Liouville theories coupled to a host of scalars.  

Going back to our strategy of axially coupling twisted chirals to the gauge multiplets of a chiral GLSM, and vice versa, a little play leads us to the very general form,
\[
{\cal L} = {\cal L}_{V}(\Phi,\S) + {\cal L}_{W}(Y,S) ~+ \int d^{2}\tt ~  \S G(Y)    ~+ \int d^{2}\t ~ S F(\Phi)+\textrm{h.c.},
\]
where ${\cal L}_{V,W}$ are the Lagrangians for standard chiral (twisted chiral) GLSMs on $V$ ($W$), $F$ and $G$ are gauge invariant analytic functions of the chiral and twisted chiral fields, repectively, and $S$ is the chiral field-strength of the gauge field in ${\cal L}_W$.  The resulting geometry has an obvious local product structure, $M \sim V \times W$, but is globally non-trivial -- this is a simple extension of the fibration structure discussed above.  One annoying feature of all such models is that any model of this form, which has trivial one-loop running of the D-term (\ie\ all the Ricci-flat manifolds), appears to be, at first blush, non-compact: it is simply impossible to build a non-trivial coupling of this form when $V$ and $W$ are both compact \CY s.  Something remains missing.

Note that the models described above evaded the ``no-go'' statement by coupling a $(2,2)$ vector minimally to chiral matter and axially to twisted chirals or vice vera.  While these models have a particularly simple presentation, they are by no means the most general $(2,2)$ models one can construct -- in particular, there are many more representations than simply chiral and twisted chiral.  In fact, as has only recently been proven\cite{Lindstrom:2005zr}, the most general off-shell $(2,2)$ NLSM can only be written by including semi-chiral multiplets anihilated by a single supercharge.  It is reasonable to ask if the same is true of GLSMs.

As it turns out, a large class of generalized geometries \cite{Hitchin:2004ut,Gualtieri:2003dx} only admit gauged linear descriptions using semi-chiral superfields.  Suppose we want to couple a $(2,2)$ gauge field to a conserved current; of necessity, that current must be either a chiral or a twisted chiral current.  However, the matter fields which appear in the current do not have to be chiral or twisted chiral, only the total current is so constrained.  This suggests a simple strategy for constructing a $(2,2)$ GLSM out of semi-chiral fields: begin with a theory of free semi-chiral fields and identify a chiral isometry of this free theory under which the semi-chiral matter fields rotate by a chiral phase.  Then, couple the associated current to a canonical $(2,2)$ gauge supermultiplet.  The result is a manifestly $(2,2)$ GLSM which, in general, does not reduce to a theory of chirals.


There are many fun $(2,2)$ torsion linear sigma models one can build, with interesting geometric and algebraic properties, but our interests in this note lie with the heterotic string, so we now turn to $(0,2)$ models, leaving a thorough discussion of the $(2,2)$ case (and the intriguing liminal $(1,2)$ case) to another publication.

\section{Non-Compact $(0,2)$ Models and the Bianchi Identity}
\setcounter{equation}{0}

Suppose we are handed a well-behaved $(0,2)$ GLSM for a vector bundle $\V_{S}$ over some happy \Ka\ manifold $S$. The FI parameters of the GLSM, $t^{a}$, parameterize some of the complexified \Ka\ moduli of $S$.  As in the $(2,2)$ cases discussed above, introducing non-trivial $H$ into this $(0,2)$ GLSM is a simple matter of promoting some subset of the FI parameters $t^{a}$ to dynamical fields $Y_{l=1...m}$ in the GLSM.  The FI coupling in a $(0,2)$ model is again a superpotential interaction, so the requisite promotion is
\[
\frac{i}{4}\int d\t^{+}~ t^{a}\U_{a} + \mathrm{h.c.}~ \to ~
\frac{i}{4}\int d\t^{+}~ N_l^{a}Y_{l}\U_{a}+\mathrm{h.c.} ~- i \int d^{2}\t ~ \overline{Y}_{l}\p_{-}Y_{l}
\]
where $y_{l} =r_{l}+i\t_{l} \in\IC^{*}$, and with $N_l^{a}\in\IZ$ to ensure single-valuedness of the action. This results in non-trivial NS-NS 3-form flux,
\[
B=N^{a}_l\t_{l} H_{a} ~~ \Rightarrow ~~ H= N^{a}_l d\t_{l} \wedge H_{a},
\]
not on $S$, but on a non-compact fibration $(\IC^{*})^{m}\to \XT \to S$, with $H$ having two legs along $S$ and one along the fibre. (Here, $H_a$ is the $a^{th}$ hyperplane class on $S$.)

This model has two major limitations.  First and foremost is the fact that the Bianchi identity is solved rather trivially: $dH=0$ by construction, since both $d\t_{l}$ and $H_{a}$ lift trivially to closed forms on the total space of the $(\IC^{*})^{m}$-fibration, $\XT$.  What we are after is an interesting solution to the Bianchi identity.  Secondly, the classical moduli space, $\XT$, is non-compact. Since the non-compactness is due to the unconstrained real part of the dynamical FI parameters, we might try to simply lift them, leaving the imaginary part dynamical as required for non-trivial $H$-flux.\footnote{Indeed, this is what happens in the Goldstein-Prokushkin construction \cite{Goldstein:2002pg}, whose compact \nK\ manifolds arise as the unit-circle sub-bundles of two $\IC^*$-bundles over a base \CY, as described further in appendix B.}  Unfortunately, this explicitly breaks $(0,2)$ supersymmetry.  In the remainder of this section we will focus on correcting the triviality of the Bianchi identity -- the thorny problem of compactification we defer to the next section.

To begin, note a curious difference from the $(2,2)$ case above.  In a $(0,2)$ gauge theory, the FI parameter does not appear in a twisted chiral superpotential -- indeed, there {\em is} no twisted chiral representation of $(0,2)$ -- but in a {\em chiral} superpotential, so the dynamical FI parameters in a $(0,2)$ theory are {\em chiral}, just like the minimally coupled scalars.  This raises an interesting possibility: since supersymmetry no longer forbids the minimal coupling of the gauge fields to the $Y_{l}$, we can couple $Y_{l}$ both axially {\it and} minimally in a completely supersymmetric fashion:
\[
{\cal L} = - \frac{1}{2}\int d^{2}\t ~ (\overline{Y}_{l} + Y_{l} + 2M_{l}^{a} V_{+a})(i\p_{-}[Y_{l} - \overline{Y}_{l}] - M_{l}^{a} V_{-a})
       + \frac{i}{4}\int d\t^{+} ~  N^{a}_lY_{l}\U_{a} +\mathrm{h.c.},
\]
where the $M_{l}^{a}$ are integers (we will discuss their quantization later).  Unfortunately, under a gauge transformation $Y_{l}\to Y_{l}-iM_{l}^{b}\Lambda_{b}$, the superpotential transforms as 
\[
\d_{\Lambda} {\cal L} =  \frac{1}{4}\int d\t^{+} ~ M_{l}^{b}N_{l}^a  \Lambda_{b}\U_{a} +\hc,
\]
which is not a total derivative, so this Lagrangian does not appear to be terribly useful.

However, this gauge variation has the familiar form of the {\em gauge anomaly} of a $(0,2)$ GLSM.  
Consider a GLSM for a holomorphic vector bundle $\V_{S}$ over a \CY\ base, $S$, built out of chiral superfields $\Phi_I$ and fermi superfields $\Gamma_m$ (see appendix A for conventions).  While the classical Lagrangian is manifestly gauge invariant, the measure generically suffers from a set of one-loop exact chiral gauge anomalies\footnote{Such gauge anomalies are strictly absent in $(2,2)$ models, where left- and right-handed fermions are paired up in $(2,2)$ chiral multiplets to give an overall non-chiral theory;  in a $(0,2)$ model, by contrast, left- and right-moving fermions transform in different \susy\ multiplets and may thus transform differently under the gauge symmetry, leading to the gauge anomaly advertised above.} of the form 
\[
\mathcal{D} [ \Phi, \G ] \stackrel{\d_{\Lambda}}{\longrightarrow} \mathcal{D} [\Phi,\G] \exp\left( -\frac{iA^{ab}}{8\pi}\int\! d^2\! y\! \left[ \int d\t^{+}   \Lambda_{b}\U_{a} + \hc \right]\right),
\]
where $A^{ab}$ is a quadratic form built out of the gauge charges 
$Q^{a}_{I}$ and $q^{a}_{m}$ of the right- and left-moving fermions,
\be
A^{ab}=\sum_{I} Q^{a}_{I}Q^{b}_{I}        ~-\sum_{m} q^{a}_{m}q^{b}_{m}.
\ee
This can be easily derived by examining the loop diagram with two external gauge bosons.  This anomaly, a familiar feature of $(0,2)$ GLSM building, has a natural geometric interpretation.  Recall that the right-handed fermions transform as sections of a sheaf $\F_V$ over the ambient toric variety $V$ which restricts over $S$ to the tangent bundle, $T_{S}$.  Meanwhile, the left-handed fermions transform as sections of a sheaf $\V_V$ which restricts to the gauge bundle, $\V_{S}$. The gauge anomaly measures 
\[
\mathcal{A} \propto ch_{2}(\F_V) - ch_{2}(\V_{V}).
\]
Since the Bianchi identity is just the restriction of $\mathcal{A}$ to $S$, the vanishing of the gauge anomaly\footnote{Note that the gauge anomaly may fail to vanish even when the classical moduli space of the GLSM has vanishing $ch_{2}$ anomaly.  For example, consider a $(0,2)$ model for an elliptic curve in $\IP^2$ with trivial left-moving bundle.  A NLSM on an elliptic curve cannot have a $ch_2$ anomaly -- nonetheless, the GLSM has a gauge anomaly.  What is going on?  The point is that the gauge anomaly computes the non-vanishing self-intersection number of the hyperplane class in $\IP^2$, an intersection which does not restrict to the hypersurface (indeed, there is no four-cohomology on $T^{2}$).  This is a somewhat familiar fact in $(0,2)$ model building: many geometries for which a NLSM analysis is perfectly consistent do not seem to admit GLSM descriptions due to uncanceled gauge anomalies.} ensures that the IR NLSM satisfies the heterotic Bianchi identity with $dH=0$.  This connection will be better explored in section \ref{secBI}.

These two effects -- the gauge variance of the classical action and the one-loop gauge anomaly -- dovetail beautifully.  Consider a GLSM for $\V_{S}\to S$ with $ch_{2}(T_{S}) \neq ch_{2}(\V_{S})$.  On its own, this model is anomalous. Now promote some subset of FI parameters to dynamical fields $Y_{l}$ with axial couplings $N^{a}_l$ and charges $M^{a}_{l}$.  Under a gauge variation, the effective action ($S_{\mathit{eff}} = \frac{1}{4\pi}\int d^2y \,{\cal L}_{\mathit{eff}}$) picks up classical terms from the axions and one-loop terms from the anomaly,
\[
\d_{\Lambda} {\cal L}_{\mathit{eff}} =  \half \int d\t^{+} ~ \left[\frac{1}{2} M_{l}^{b}N^{a}_l - Q^{a}_{I}Q^{b}_{I}  + q^{a}_{m}q^{b}_{m}\right]  \Lambda_{b}\U_{a} + \hc \, .
\]
Thus, for every solution of the Diophantine equation 
\be\label{eq:anomaly}
\frac{1}{2}\sum_l M_{l}^{b}N^{a}_l  = \sum_{I} Q^{a}_{I}Q^{b}_{I}  - \sum_{m} q^{a}_{m}q^{b}_{m}
\ee
we have a non-anomalous $(0,2)$ quantum field theory. Since the superpotential of this $(0,2)$ theory is not renormalized beyond one loop in perturbation theory, and since the anomaly is one-loop exact, the path integral remains gauge invariant to all orders in perturbation theory.\footnote{We will discuss non-perturbative effects below.}  Note that the $ch_{2}$ anomaly in the NLSM is also one-loop exact.  We shall refer to a $(0,2)$ GLSM which implements the above cancellation mechanism as a \emph{torsion linear sigma model} (TLSM).

Notice what has happened.  First, we have replaced the \Ka\ geometry $S$ with a \nK\ $(\IC^{*})^m$-fibration $\XT$ over $S$ such that the curvature 2-forms of the $(\IC^{*})^m$-fibration are trivial in $H^{2}(\XT,\IZ)$, the cohomology of the total space.   It is important to distinguish $ch_{2}(T_{S})-ch_{2}(\V_{S})$, the anomaly on $S$, from the very different quantity  $ch_{2}(T_{\XT})-ch_{2}(\V_{\XT})$, the anomaly on the $(\IC^{*})^{m}$-fibration $\XT$ {\em over} $S$.   At the end of the day, the physical Bianchi identity lives on $\XT$ and says that $dH=ch_{2}(T_{\XT})-ch_{2}(\V_{\XT})$, so in cohomology on $\XT$, $ch_{2}(T_{\XT})=ch_{2}(\V_{\XT})$.  However, since $\XT$ is a non-trivial fibration over $S$, cohomology classes do not trivially lift, or descend (think about the Hopf map).  The upshot it that Bianchi identity does {\em not} imply that $ch_{2}(T_{S})=ch_{2}(\V_{S})$, even in cohomology.  However, the 3-form flux $H=N_l^{a} d\t_{l}\wedge H_{a}$ on the the total space, $\XT$, was constructed precisely so as to solve the Bianchi identity when pushed down the fibres -- this is what led us to introduce the gauge-variant axial coupling in the first place.

This graceful mechanism of anomaly cancellation, a one-loop gauge anomaly canceling  the gauge variation of an axionic coupling in the classical Lagrangian, is simply a 2d avatar of the Green-Schwarz anomaly in the target space.

\section{Compact $(0,2)$ Models and the Torsion Multiplet}
\setcounter{equation}{0}

Let us summarize the story so far.  We begin with a conventional $(0,2)$ GLSM for a Calabi-Yau $S$ equipped with a generic holomorphic bundle $\V_{S}$.  The $ch_{2}(T_{S})\neq ch_{2}(\V_{S})$ anomaly of the associated NLSM is realized in the GLSM as a gauge anomaly.  To cancel the gauge anomaly, we promote some of the FI parameters to dynamical axions carrying charges chosen such that the gauge variation of the classical action cancels the one-loop gauge anomaly in a 2d version of the Green-Schwarz mechanism.  The IR geometry of the resulting non-anomalous $(0,2)$ GLSM is a non-compact $(\IC^{*})^{m}$-fibration $\XT$ over $S$,
\bea
(\IC^{*})^{m} \longrightarrow &\XT& \non\\
&\downarrow& \non\\
&S&, \non
\eea
where the curvature two-forms of the $\IC^{*}$-bundles are $M_l^{a} H_{a}|_S\in H^{2}(S,\IZ)$.  Threading this geometry is a non-trivial NS-NS 3-form flux, $H=N_l^{a}d\t_{l}\wedge H_{a}$, which satisfies the Bianchi identity non-trivially.  For simplicity of presentation, we will focus on the special cases $S=K3$ or $T^4$ with $m$=2; the generalization to higher dimension and other geometries is straightforward.  

Not coincidentally, this is enticingly close to the compact \FY\ geometry\footnote{A review of the \FY\ geometry is given in appendix B.} -- all we need to do is restrict to the $T^{2}$ sub-bundle of the $(\IC^{*})^{2}$ bundle by lifting the real direction along each $\IC^{*}$ fibre.  What could be easier?

\subsection{Decoupling of Radial Fields}

In fact, this turns out to be rather non-trivial.  The issue is supersymmetry.  The target space of any sigma model with a linearly realized $\N=2$ is a complex manifold, and the specific presentation of the $\N=2$ corresponds to a specific choice of complex structure.  Under the particular $\N=2$ respected by our GLSM, the real directions along the $\IC^{*}$ fibre, $r_{l}$, are paired with the $S^{1}$ angles, $\t_{l}$, so removing only the radial coordinates would explicitly break our $(0,2)$ \susy\ to an all-but-useless $(0,1)$ subgroup (which we are not allowed to lose since this $(0,1)$ will be gauged when we couple our matter theory to heterotic worldsheet supergravity).  The situation appears to be grim.

To reassure ourselves that there \emph{should} be a $(0,2)$ on the $T^{2}$ sub-bundle, note that
\[
(\IC^{*})^{2} = \IC \times T^{2}
\]
if the coordinates $y_{l}=r_{l}+i\t_{l}$ on $(\IC^{*})^{2}$ are reorganized into the coordinates 
$r=r_{1}+ir_{2}$ on $\IC$ and $\theta=\t_{1}+i\t_{2}$ on $T^{2}$.  The IR geometry thus must admit an $\N=2$ corresponding to this choice of complex structure, pairing the two angles into one supermultiplet and the two lines into another.  Unfortunately, an extensive search for such an $\N=2$ in our UV gauge theory quashes our high expectations.

Let's explore this apparent failure more explicitly.  The relevant terms in the action are, in 
components,
\bea
{\cal L} &=& {\cal L}_{K3} 
- k_l^{2}(\p r_{l})^{2}
- k_l^{2}(\p \t_{l} +M_{l}^{a}v_{a})^{2}
+2ik_l^{2}\bar{\chi}_{l}\p_{-}\chi_{l} +2N^{a}_l\t_{l}v_{+-a} \non\\
&& + \left(2k_l^{2}M^{a}_l-N^{a}_l\right) \left[r_{l}D_{a} + \frac{i}{\sqrt{2}}\chi_{l}\lambda_{a}  \right]+\ldots \,. \non
\eea
Meanwhile, under the linearly realized $(0,2)$ \susy\ 
\be{\label{sssvar}
\d_{\e} \lambda_{a} = i\e(D_a+2iv_{+-a}) 
}\ee
Now, suppose we attempt reorganize the $Y_{l}$ superfields into superfields that respect the $\IC \times T^{2}$ complex structure: $R\sim r_1 + ir_2 + \ldots$, $\Theta\sim \t_1 + i\t_2+\ldots$.  
The problem is that the variation of $\lambda_a$ yields terms of the form $\epsilon\chi_l D_a$ and $\epsilon\chi_l v_{+-a}$.  The only way to cancel these terms is for the variation of both $r_l$ and $\t_l$ to include terms of the form $\epsilon\chi_l$.  This makes it appear impossible to split $r_l$ and $\t_l$ into two separate supermultiplets for generic charges.

The key word here is ``generic''.  Note that our troublesome terms are both proportional to $(2k_{l}^{2}M_l^{a}-N_l^{a})$, where $M_l^{a},N_l^{a}\in\IZ$ and $k_{l}\in\IR$.  If we fix $k_{l}$ and $N_l^{a}$ so that $N_l^{a}=2k_{l}^{2}M_l^{a}$, these terms disappear from the action!  Repeating our analysis, we find that there \emph{is} a $(0,2)$ \susy\ with exactly the desired properties:
\[
R= (r_1- ir_2) + i\sqrt{2}\theta^{+}(\chi^{I}_{1} + i \chi^{I}_{2}) + ...
~~~~~~~~
\T= (\t_1 +i\t_2) + \sqrt{2}\theta^{+}(\chi^{R}_{1} - i \chi^{R}_{2}) + ...
\]
where $R$ and $I$ superscripts refer to the real and imaginary parts of the fermions, respectively.  
In fact, the $R$-multiplet is free and {\it entirely decouples!}  What's more, since $k_{l}$, which measures the radius of the $T^{2}$ in string units, is fixed in terms of two integers, the volume of the fibre is quantized in terms of the torsion flux, just as it is in Fu-Yau.\footnote{Since $\int d^2y \, v_{+-a} \in \pi\IZ$, $\t_{l}$ is automatically periodic, $\t_{l}\sim\t_{l} + 2\pi L_{l}$, such that $N_{l}^{a}L_{l} \in 2\mathbb{Z}$.  Fixing $N^{a}_{l} = 2k_{l}^2 M^{a}_{l}$ then implies that $k_{l}^2 M^{a}_{l}L_{l} = n_{a} \in \mathbb{Z}$, so $M^{a}_{l}$ is quantized in terms of $k_{l}$ and $L_{l}$.  Meanwhile, the anomaly cancellation condition implies that that $\frac{n_{a}^2}{k_{l}^2 L_{l}^2}$ should be an integer, since the $Q_I$ and $q_m$ are integers.  Since the physical radius of the $S^1$ is $k_{l}L_{l}$, this means that the radius is quantized as claimed.  For the rest of this paper, we will work with $k_{l}=L_{l}=1$ for simplicity.}

Life is now sweet and easy.  Based on the above, we define
\be
\begin{array}{lll} 
      \t=\t_{1}+i\t_{2} 	~~& \chi=\chi^{R}_{1}-i\chi^{R}_{2}	~~& N^{a}=2k^2M^a=2k^{2}(M_1^{a}+iM_2^{a}) \\
      r=r_{1}-ir_{2} 	~~& \tilde{\chi}=i\chi^{I}_{1}-\chi^{I}_{2}~~& \nabla_{\pm}\t = \p_{\pm}\t +M^{a} v_{\pm a},
\end{array}
\ee
which transform under $\N=2$ \susy\ as
\be
\begin{array}{ll} 
	\d_{\e}\t = -\sqrt{2}\e\chi 		~~~~&	\d_{\e}\chi = 2\sqrt{2}i\bar{\e} ~\nabla_{+}\t  \\
	\d_{\e} r = -\sqrt{2}\e\tilde{\chi}	~~~~&	\d_{\e}\tilde{\chi} = 2\sqrt{2}i\bar{\e}~\p_{+}r. 
\end{array}
\ee
In these coordinates, the action reduces to
\bea
{\cal L} &=& {\cal L}_{K3} 
+ 2\nabla_{+}\bar{\t}\nabla_{-}\t + 2\nabla_{+}\t\nabla_{-}\bar{\t} + 2i\bar{\chi}\p_{-}\chi + 2(N^{a}\bar{\t} +\overline{N}^{a}\t)v_{+-a} \non\\
&& - 2 |\p r|^{2}+ 2i\bar{\tilde{\chi}}\p_{-}\tilde{\chi}.
\eea
We may now drop the radial supermultiplet $R=r+\sqrt{2}\t^{+}\tilde{\chi}-2i\t^+\bar{\t}^+\p_+ r$, as it is entirely decoupled.

It is important to verify that the truncated Lagrangian is invariant under the $(0,2)$ \susy\ defined above.  However, $\d_{susy}^{2}=\d_{gauge}$ in WZ gauge, so the gauge variance of the classical action rears its stupefying head and some care is required.  Under a supersymmetry transformation, the classical action transforms non-trivially,
\[
\d_{\e}{\cal L} = 2(M^{a}\bar{M}^{b} +\bar{M}^{a}M^{b})v_{+b}(i\e\bar{\lambda}_a + i\bar{\e}\lambda_a).
\]
This is not a disaster because the gauge transformation needed to return us to WZ gauge (which we have been using throughout), $\a_{a}=-4i\t^{+}\bar{\e}v_{+a}$, induces a shift in the effective action from the anomalous measure:
\[
\delta_{WZ}{\cal D} [ \Phi,\Gamma ]={\cal D} [ \Phi,\Gamma ] \exp\!\left\{ \frac{A^{ab}}{\pi} \int d^2y \, v_{+a}(\e\bar{\lambda}_{b}+i\bar{\e}\lambda_{b})\right\}.
\]
Anomaly cancellation ensures that this cancels the \susy\ variation of the action.

At this point, we can play various games to simplify the presentation of the theory.  For example, we can build a superfield out of the $T^{2}$ multiplet,
$$
\Theta = \t + \sqrt{2}\t^{+}\chi-2i\t^{+}\bar{\t}^{+}\nabla_{+}\t.
$$
This looks a lot more convenient than it actually is.  While it has the usual field content, this is {\em not} a standard chiral multiplet: the gauging is complex, with both real and imaginary components of $\t$ shifting under gauge transformations.  Since no other superfield transforms in the same strange way, gauge multiplet included, it is extremely hard to build gauge covariant or invariant operators out of $\Theta$.  In fact, the only gauge-invariant dressed field is $(\p_{-}\T+\half M^{a}V_{-a})$.  Meanwhile, the only chiral operator we can build out of $\T$ is $(\T+iM^{a}V_{+a})$, which we cannot add to the superpotential in a gauge invariant fashion.  Indeed, it is impossible to build a supersymmetric and gauge invariant action for this multiplet since the supersymmetry variation of the kinetic terms cancels against the variation of the axial superpotential.  To emphasize its peculiar role, we call $\Theta$ a \emph{torsion multiplet}.

\subsection{The IR Geometry}\label{sec:5}

Setting $N_l^{a}=2k^{2}M_l^{a}$ has decoupled the $R$ multiplet, leaving us with a non-K\"{a}hler $T^{2}$ sub-bundle $X\subset {\XT}$ with torsionful $SU(3)$-structure induced from $\XT$.   In other words, the semi-classical IR geometry of our TLSM is a compact holomorphic $T^{2}$ fibration $X$ over a \CY\ $S$, endowed with a Hermitian metric, a stable holomorphic sheaf $\V_{X}=\pi^{*}\V_{S}$ pulled back from $S$, and an NS-NS 3-form $H$ satisfying the Bianchi identity on $\V_{X}\to X$.   Moreover, the radii of the $T^{2}$ fibres are fixed to discrete values in terms of the integral curvatures of the $T^{2}$-bundles, given as integer classes on the base $K3$.  Up to uninteresting changes of coordinates, this is the \FY\ construction.

It is revealing to derive this IR geometry explicitly from the final TLSM.  Let's begin by writing out the component Lagrangian in all its majesty.  To simplify our lives, we will call all the chiral multiplets $\phi_{I}$ whether their charges are positive, negative, or zero, and leave all obvious sums implicit.  This is easy to unpack when we focus on specific models. After integrating out the auxillary fields, the kinetic terms are,
\bea
{\cal L}_{kin} &=& 
    -|(\p+iQ^{a}_{I}v_{a}) \phi_{I}|^{2}   
 + 2i\psibar_{I}(\p_{-}+iQ^{a}_{I}v_{-a})\psi_{I} \non\\
&& +  4(\p_{+} \t_{l} + M_{l}^{a}v_{+a})(\p_{-}\t_{l} + M_{l}^{a}v_{-a})   + 4M_{l}^{a}\t_{l}v_{+-a} + 2i\chibar_{l}\p_{-}\chi_{l}    \non\\
&& +  2i\gbar_{m}(\p_{+}+iq^{a}_{m}v_{+a})\g_{m}
 +    \frac{2}{e_{a}^{2}} \left[   ( v_{+-a} )^2  + i\lb_{a}\p_{+}\lambda_{a}  \right],   \non
\eea
and the scalar potential is
\bea
U = \sum_{m}\left( |E_{m}|^{2} + |J^{m}|^{2}\right)  + \sum_{a} \frac{e_{a}^{2}}{2}\left(\sum_{I} Q_I^a |\phi_{I}|^{2} - r^{a}\right)^2
\eea
where $\overline{\CD}_{+}\Gamma_{m}=\sqrt{2}E_{m}(\Phi)$ and $J^{m}(\Phi)$ is a $(0,2)$ superpotential satisfying $\sum_{m}E_{m}J^{m}=0$.  For completeness, the Yukawa terms are
\bea
{\cal L}_{Yuk} = - \sqrt{2}iQ_I^a \lambda_{a} \psi_{I} \bar{\phi}_{I}  -  \gbar_{m}\psi_{I} \frac{\p E_{m}}{\p \phi_{I}}  - \g_{m}\psi_{I} \frac{\p J^{m}}{\p \phi_{I}} +\hc\, .
\eea

As in the case of $(2,2)$ GLSMs on \Ka\ geometries, the Hermitian geometry of the Higgs branch of our TLSM may be computed by integrating out the massive vectors and scalars in the gauge theory to derive a Born-Oppenheimer effective action on the classical moduli space.  However, since the classical action of our TLSM is not gauge invariant, the story is slightly more subtle than usual.  

Suppose, for example, that we simply integrate out the massive vector as usual -- let us work in polar variables where $\phi_I = \rho_I e^{i\vp_I}$.  This replaces the gauge connection $v_{\mu}$ with a non-trivial implicit connection $v_{\mu}(\rho_{I},\vp_{I},\t_l,\ldots)$ on the classical moduli space.  The chiral fermion content then leads to an anomaly in the resulting non-linear sigma model -- an anomaly which cancels against the classical variation of the action due to the torsion multiplet.  This presentation has the advantage of making the role of the anomalous gauge transformation in the NLSM manifest, but it complicates the computation of the effective metric.

Alternatively, we can take a lesson from Fujikawa and change coordinates in field space to work with uncharged fermions {\em before} integrating out the massive vector\cite{Fujikawa:1979ay,Fujikawa:1980eg}.  The Jacobian of this field redefinition introduces a gauge variant operator to the action whose gauge variation cancels against that of the classical torsion terms, leaving the action gauge invariant.  We can then integrate out the massive vector and massive scalars to compute the effective metric on moduli space.

Let's take the second approach and change variables to gauge invariant fermions.  
For each right-moving fermion $\psi_{I}$, there is a natural choice of uncharged dressed fermion $\psit_{I}=e^{-i\vp_{I}}\psi_{I}$; for the left-movers, there is generically no model-independent choice, so we choose an arbitrary linear combination $\hat{\vp}_{m}=l_{m}^{I}\vp_{I}$ of phases with the correct charges to make the dressed fermion $\gt_{m}=e^{-i\hat{\vp}_{m}}\g_{m}$ gauge neutral, \ie\ such that $\delta_{\a} \hat{\vp}_{m} = -q_{m}^{a}\a_{a}$.  The Jacobian for this change of variables shifts the action by a simple term 
\[
{\cal L}_{Jac} =  -4 \w^{a}  v_{+-a}, \qquad  \w^{a} \equiv Q^{a}_{I}\vp_{I}-q_{m}^{a}\hat{\vp}_{m} \equiv T_I^a\vp_I ,
\]
whose gauge variation is just the familiar anomaly,
\[
\d_{\alpha} {\cal L}_{Jac} = 4 \left( Q^{a}_{I}Q^{b}_{I} - q_{m}^{a}q_{m}^{b} \right) \alpha_{a} v_{+-b}.
\]
The total axial coupling is thus
\[
{\cal L}_{axial} =  4 \left( M_{l}^{a}\t_{l} - \w^{a} \right)  v_{+-a},
\]
which is gauge invariant by construction.  The typical next step is to fix a gauge.  However, since the Faddeev-Popov measure for the simplest gauge choice, $\t_l=0$, is trivial, it is just as easy to work in gauge unfixed presentation; the decoupled longitudinal mode will simply cancel the volume of the gauge group in the path integral.

With the action and measure now both independently gauge invariant, we can consistently integrate out the massive vector.  Since the action is quadratic in the vector, this is straightforward.  Solving the classical EOM for the two components of our massive vector, and splitting them into fermionic and bosonic components, yields
\bea
v_{-a}  &=& (\DI)^{ab} \lp \half \gtb_{m}\gt_{m}q^{b}_{m} - \rho^{2}_{I}\p_{-}\vp_{I}Q^{b}_{I} + \p_{-}\w^{b} - 2M_{l}^{b} \p_{-}\t_{l} \rp  =  v^F_{-a} + v^B_{-a}   \non \\
v_{+a} &=& (\DI)^{ab} \lp \half \psitb_{I}\psit_{I}Q^{b}_{I} - \rho^{2}_{I}\p_{+}\vp_{I}Q^{b}_{I} - \p_{+}\w^{b} \rp   =   v^F_{+a} + v^B_{+a}  \non 
\eea
where we define
\be
\Delta^{ab} \equiv \rho_I^2 Q_I^a Q_I^b + M_l^a M_l^b \equiv \Delta_{Q}+\Delta_{M}, \non
\ee
which is naturally symmetric in the gauge indices.  It is easy to check that both components of $v$ transform covariantly under gauge transformations.

Thus prepared, we are finally ready to compute the effective metric on the Higgs branch.  After a tedious but miserable calculation, the bosonic effective action reduces to
\bea
{\cal L}^B_{kin}
&=& 4\p_+\rho_I\p_-\rho_I + 4\p_+\vp_I \p_-\vp_J \left[ \rho_I^2 \delta_{IJ} - \rho_I^2 \rho_J^2(\DI)_{ab}Q^a_I Q^b_J \right] + 4(\DI)_{ab}\p_+\w^a\p_-\w^b      \non \\
&& + 4\p_+\t_l\p_-\t_l
- 8(\DI)_{ab}\left(\rho_I^2\p_+\vp_I Q_I^a + \p_+\w^a\right)\p_-\t_l M_l^b 
- 8(\DI)_{ab} \rho_I^2 Q_I^a\p_{[+}\w^b\p_{-]}\vp_I \, ,  \non
\eea
and the fermionic effective action to
\bea
{\cal L}^F_{kin} 
&=&
2i\psitb_{I}(\p_{-}+iQ^{a}_{I}v^B_{-a}+i\p_-\vp_I)\psit_{I} + 2i\chibar_{l}\p_{-}\chi_{l} + 2i\gtb_{m}(\p_{+}+iq^{a}_{m}v^B_{+a}+i\p_+\hat{\vp}_m)\gt_{m}  \non \\
& & - (\DI)_{ab}\psitb_I\psit_I\gtb_m\gt_m Q_I^a q_m^b \, , \non
\eea
where $A_{[+}B_{-]}\equiv \half (A_+B_- - A_-B_+)$, ~$A_{(+}B_{-)} = \half(A_+B_- + A_-B_+)$.  
We will also find it useful to define $\DI_2 \equiv \DI - \DI_Q = -\DI_Q \Delta_M \DI $, and to make a habit of suppressing gauge indices, representing them instead by matrix multiplication.   

Since one of the features we would like to make manifest is the natural complex structure on the total space $X$, it is natural to return to complex variables $\phi_I$ and $\t$, as well as $M^a \equiv M_1^a + iM_2^a$.  It is also natural to split the Lagrangian into terms symmetric and anti-symmetric in the derivatives, corresponding to the pullback to the worldsheet of the metric and $B$-field, respectively.  The symmetric terms we will refer to as $ds^2$, where we will also use the shorthand
\be
dA dB \equiv \partial_{(+}A\partial_{-)}B  \qquad   dA\wedge dB \equiv \p_{[+}A\p_{-]}B ,  \non
\ee
remembering that the ``differentials'' $dA$ and $dB$ are symmeterized without the $\wedge$.

Using these conventions and the definition of $\DI_2$, we can easily factor out the usual kinetic terms for the ambient variety $V$:
\be
ds^2_V = 4|d\phi_I|^2 - 4(\phib_I d\phi_I)(\phi_J d\phib_J)Q_I^T \DI_Q Q_J .   \non
\ee
The metric can then be written as
\bea
ds^2 &=& ds_V^2 - 4|\phi_I|^2|\phi_J|^2 (d\ln\phib_I d\ln\phi_J) Q_I^T\DI_2 Q_J
- \left[ d\ln\frac{\phi_I}{\phib_I}\right]\left[ d\ln\frac{\phi_J}{\phib_J}\right] T_I^T \DI T_J    \non \\
&& + 4|d\t|^2 + 2i \left[ d\ln\frac{\phi_I}{\phib_I}\right]\left( |\phi_I|^2 Q_I^T + T_I^T\right)\DI\left(Md\bar{\t}+\bar{M}d\t\right)
\eea
where we have used $dr^a = \sum_I Q_I^a (\phi_I d\bar{\phi_I} + \bar{\phi}_I d\phi_I) = 0$ to simplify the expression.  Working patchwise on $V$ makes the geometry somewhat more transparent.  We can cover $V$ by patches on which $s$ of the homogeneous coordinates, say $\phi_{\sigma=N+1,\ldots,N+s}$, are nonzero and for which $Q_\sigma^a$ is an invertible $s\times s$ matrix.  We can then define gauge invariant coordinates on each patch,
\be
z_A \equiv \phi_A \prod_{\sigma=N+1}^{N+s} \phi_\sigma^{-\QI^\sigma_a Q^a_A} , \qquad   \zeta \equiv \t + i\QI^\sigma_a M^a \ln\phi_\sigma,   \non
\ee
where $A=1,\ldots,N$.

All of these coordinates transform holomorphically as we move from one patch of $V$ to another.  Furthermore, from the gauge variant coordinates it is clear that there are no fixed points of the $T^2$ action (complex shifts of $\t$).  Thus, as long as $S\subset V$ is smooth, our construction will yield a principal holomorphic $T^2$ bundle over $S$ {\it\`a la} Goldstein and Prokushkin (see appendix B).  In these manifestly holomorphic coordinates, the metric can be written in Hermitian form,
\bean
ds_H^2 &=& ds_V^2 + 4\left| d\zeta - \frac{iM^T}{2}\left( \p P - \DI(Q_A |\phi_A|^2+T_A) d\ln z_A \right)\right|^2   \\
&& + \left( \p P^T \Delta_M + d\ln z_A \, T_A^T \right) \left( 2\DI - \DI M\bar{M}^T \DI \right)\left( \Delta_M \bar{\p} P + T_B\, d\ln\bar{z}_B \right)
\eean
where $P_a \equiv \sum_\sigma \QI^\sigma_a \ln|\phi_\sigma|^2$ and
\bean
ds_V^2 &=& 4|\phi_A|^2|d\ln z_A|^2 - 4\left(|\phi_A|^2 Q^T_A\DI_Q Q_B |\phi_B|^2\right)\left[d\ln z_A\right] \left[d\ln\zb_B\right] 
\eean
is the analog of the Fubini-Study metric for $V$ (and reduces to it in the case of $\mathbb{P}^N$).  
A similarly tedious but straightforward computation gives the resulting $B$-field,
\bean
B &=& 2i\left[ d\ln\frac{z_A}{\zb_A}\right] \wedge (|\phi_A|^2 Q_A^T + T_A^T) \DI \left[ M d\bar{\zeta} + \bar{M} d\zeta\right]    \non \\
&& - 2(|\phi_A|^2 Q_A^T\DI T_B)\left[ d\ln\frac{z_A}{\zb_A}\right]\wedge\left[d\ln\frac{z_B}{\zb_B}\right]     \non \\
&& - (\bar{M}^a M^T - M^a\bar{M}^T)\DI(Q_A|\phi_A|^2+T_A)\left[ d\ln\frac{z_A}{\zb_A} \right]\wedge dP_a .
\eean
We thus have a manifestly Hermitian metric on a smooth principal holomorphic $T^{2}$-bundle over $S$, with non-vanishing $H$ threading the total space.  This is precisely the geometry we were expecting to find.

\subsection{The Bianchi Identity}\label{secBI}

As we sketched in section 3, the one-loop exact spacetime Bianchi identity is realized in the TLSM by the one-loop exact gauge anomaly.  However, the gauge anomaly is independent of the superpotential and thus naturally lives on the ambient toric variety $V$, while the Bianchi identity lives on the space $X$, so the connection between the Bianchi identity and the gauge anomaly requires some work to explicate.

Their relationship is most transparent when the Bianchi identity is pushed down to the base, $S$.  In the Fu-Yau case, it has been shown on purely geometric grounds that  \cite{Fu:2006vj, Becker:2006et}
\[
dH=\pi^*\left(\omega\wedge *_{\!S} \overline{\omega}\right) +\ldots, \qquad 
ch(T_X)=\pi^*\left(ch(T_S)\right)+\ldots\, ,
\]
where $\w=\w_1+i\w_2$ is the anti-self-dual\footnote{Strictly speaking, there can also be a self-dual $(2,0)$ $\omega$-form, but it is automatically absent in the TLSM construction.} (1,1) curvature form of the $T^2$ bundle, and the omitted terms are all exact forms on $S$ and thus vanish in cohomology on the base. Meanwhile, by construction, $ch(\V_{X})=\pi^*\left(ch(\V_{S})\right)$, so the Bianchi identity reduces to a simple equation in the cohomology of $S$:
\be\label{eq:BIonS}
\omega\wedge *_{\!S} \overline{\omega} = -\w_1^2 - \w_2^2 = 2ch_{2}(\V_S) - 2ch_{2}(T_{S}) .
\ee
All the quantities in this equation can now be written in terms of the defining charges of the TLSM.  The second Chern characters can be calculated from the short exact sequences \C{eq:seq1} and \C{eq:seq2} to be
\bean
ch_2(T_S) &=& \frac{1}{2}\sum_{a,b} \left(\sum_i Q_i^a Q_i^b - d^a d^b\right)\left.\left(H_a \wedge H_b\right)\right|_S,\\
ch_2(\V_S) &=& \frac{1}{2}\sum_{a,b} \left(\sum_m q_m^a q_m^b- m^a m^b\right)\left.\left(H_a \wedge H_b\right)\right|_S,
\eean
Meanwhile, the curvature $\w$ of the $T^{2}$ fibration can be expressed as $\w= (M_1^{a}+iM_2^{a}) H_a$, so the Bianchi identity pushes down to $S$ to give
\be
\label{eq:sBI}
\sum_{a,b} \left( M^{(a}\bar{M}^{b)} 
- \sum_i Q_i^a Q_i^b + d^a d^b
+ \sum_m q_m^a q_m^b - m^a m^b 
\right) \left.\left(H_a \wedge H_b\right)\right|_S = 0 .
\ee
This is precisely the condition for the cancelation of the gauge anomaly of the TLSM.

\subsection{Ruling Out $T^{4}$}

The case $S=T^{4}$ provides a revealing test case for our construction.  Since $T_{T^{4}}$
is (utterly) trivial, the Bianchi identity takes a particularly simple form -- in fact, it is so simple that there are no non-trivial solutions \cite{Becker:2006et}. This can be seen by integrating \C{eq:HBI} over the base using the restricted forms of $dH$ and $[ch(\V_X)\!-\!ch(T_X)]$ given in the previous section. Since $F_S$ -- the curvature of the bundle $\V_S$ -- is anti-Hermitian and anti-self-dual, and since $ch_2(T_{T^{4}})=0$, the right-hand side of \C{eq:BIonS} is non-positive for $S=T^4$ while the left-hand side is manifestly non-negative for anti-self-dual $\omega$ (and only $0$ when $\omega$ is exact).  We would like to see this directly in the TLSM, or at least in a specific example.

To this end, we build the base $S=T^4$ as the product of two $T^2 \subset \IP^2 $, but with $H$-flux lacing both factors.  This ensures that any 4-form on the base must be proportional to $\left.H_1 \wedge H_2\right|_S$, where $H_1$ and $H_2$ are the hyperplane classes of the two $\IP^2$s (the restrictions of $H_i^2$ vanish trivially).  Since the Hodge star on $T^4$ acts as
\[
*_{S}H_1=H_2\qquad *_{S}H_2=H_1,
\]
$(H_1-H_2)$ is the only anti-self-dual 2-form on $T^4$ constructed from hyperplane classes. Since the Fu-Yau construction requires $\w$ be anti-self-dual, we must have $\omega=M(H_1-H_2)$. Two further conditions apply: (1) for our embedding of $T^4$, none of the coordinate fields are charged under both $U(1)$s, and so $d^1d^2=Q_i^1 Q_i^2=0$; and (2) the condition that $c_1(\V_S)=0$ translates into $m^a =\sum_m q_m^a$. Plugging this into \C{eq:sBI}, only the $H_1 \wedge H_2$ cross-term does not vanish upon restriction to $S$ and we find
\be
\sum_{m \neq n} q_m^1 q_n^2 = -|M|^2.
\ee
But for the gauge bundle to be stable, all charges must satisfy $q^a_m \geq 0$ \cite{Distler:1993mk}, in which case the equation has no solution unless $M=0$.   
We conclude that our TLSM does not allow us to build a non-trivial $T^2$-bundle over this $T^4$-base, in agreement with the the supergravity result.

\subsection{Global Anomalies}

Of course, vanishing of the gauge anomaly and satisfaction of the Bianchi identity are not sufficient to ensure that the TLSM flows to a consistent vacuum of the heterotic string.  In order to couple to worldsheet supergravity, our theory must flow to a superconformal fixed point which admits a chiral GSO projection.  This in turn requires \cite{Distler:1993mk,Distler:1995mi} the existence of a non-anomalous right-moving $U(1)$ $R$-current, $J_R$, and a non-anomalous left-moving flavor symmetry, $J_L$, leading to additional constraints on allowed charges beyond quantum gauge invariance.  
The relevant anomalies are thus the various mixed gauge-global and global-global anomalies; consistency of the gauge theory requires that they cancel.

Let's start with the $R$-current.  $R$-invariance of the $\U\Theta$ terms in the superpotential require $\Theta$ to be an $R$-scalar, though it may carry a shift-charge under $R$-symmetry.  This implies that the fermion $\chi$ in $\Theta$ carries $R$-charge $+1$.  However, since $\chi$ is gauge neutral, it does not contribute to the mixed gauge-$R$ anomaly.  Since the chiral superfields $\Phi_i$ typically appear in quasi-homogeneous polynomials in the superpotential $\Gamma_0 G(\Phi_i)$ (see appendix A), it is most natural to assign them $R$-charges proportional to their gauge charges $rQ_i$ -- this also fixes the $R$-charge of $\Gamma_0$ to $-rd-1$.  Then one has the fermi supermultiplets $\Gamma_m$ appearing in the superpotential via $\Phi_0 \Gamma_m J^m(\Phi_i)$, restricting charge assignments for $\Phi_0$ and $\Gamma_m$ to be $p-rm$ and $rq_m-p-1$, respectively.  This additional shift of $p$ is a freedom not available to us in $(2,2)$ models.

The anomaly in the left-moving flavor symmetry can be treated similarly.  For example, by setting the flavor charge of each field proportional to its gauge charge, and assigning to $\Theta$ an anomalous shift-charge under the flavor $U(1)$, vanishing of the gauge anomaly ensures the non-anomaly of the left-moving flavor symmetry. Note that the contribution of the torsion multiplet to the currents $J_L$, $J_R$, and $J_{gauge}$, is of the form $J_{\Theta}\sim \p\t$, so its contributions to the anomalies actually come from tree-diagrams rather than loops.  

Two final anomaly relations are important.  First, for $J_R$ and $J_L$ to be purely right- and left-moving, their mixed anomaly must also vanish, giving one integer constraint.
Finally, the $J_{R}J_{R}$ OPE measures the conformal anomaly, which must be equal to 9, giving one last integer equation on the charges.
In the typical model of interest, there are many more fields than equations, making it easy to satisfy these constraints.

\subsection{Caveat Emptor: Spacetime vs. Wordsheet Constraints}

One very important elision in the above is distinguishing which conditions on the charges are required on {\it a priori} 2d grounds, and which derive from spacetime arguments.  
For example, in a $(2,2)$ model the running of the D-term is equivalent to the $R$-anomaly, which in turn is equivalent to the vanishing first Chern class of the IR geometry, $c_1(T_S)$.  However, in a $(0,2)$ model these three effects are decoupled.

The running of $t$ is decoupled because we can always add a pair of massive spectators to the theory -- a chiral and a fermi superfield -- whose contributions to all gauge and global anomalies vanish, but whose gauge charges can be chosen to limit the running of $t$ to a finite shift \cite{Distler:1993mk,Distler:1995mi}, something not possible in more familiar $(2,2)$ models.  Meanwhile, the chiral content of the theory yields enough freedom in assigning $R$-charges that the $R$-anomaly is decoupled from $c_1(T_S)=0$.  Similarly, the conditions that $c_{1}(\V_{S})$=0, that $\w$ be anti-self-dual, and that $\V_{S}$ be stable, are all required to ensure spacetime supersymmetry in the supergravity construction of the \FY\ compactification but do not appear as necessary constraints for the consistency of our 2d gauge theories.

A natural guess is that ensuring spacetime supersymmetry of the massless modes of our theory requires the imposition of these constraints on the charges and fields in the TLSM.  Checking this requires a more detailed discussion of the exact spectrum of our models than we have presented in this note; for now we will simply impose these conditions, as is often done in $(0,2)$ models, because we can and because doing so matches us precisely onto the \FY\ construction.  We will return to this question in a sequel.

\section{The Conformal Limit}
\setcounter{equation}{0}

So far, we have shown that our compact $(0,2)$ TLSMs exist as non-anomalous, 2d $\N=2$ quantum field theories which have Fu-Yau-type geometries as their one-loop classical moduli spaces. These are principal holomorphic $T^{2}$-bundles over Calabi-Yaus with torsionful $G$-structures which non-trivially satisfy the Green-Schwarz anomaly constraints.  However, since Fu-Yau geometries are necessarily finite radius and generally contain small-volume cycles, the semi-classical geometric analysis is not obviously reliable.  What we would like to argue is that the IR conformal fixed points to which these massive TLSMs flow should be taken to {\em define} the \FY\ CFT.  For this to make sense, however, we must demonstrate that these TLSMs in fact flow to non-trivial CFTs in the IR.

This will take some work.  The first step is to observe that the superpotential in a $(0,2)$ model is one-loop exact, so the vacuum is not destabilized at any order in perturbation theory; the concern is thus worldsheet instantons. It has long been understood that the perturbative moduli spaces of generic $(0,2)$ models are lifted by instanton effects \cite{Dine:1986zy, Dine:1987bq}.  It has more recently been understood that $(0,2)$ GLSMs on K\"{a}hler targets with arbitrary (not necessarily linear) stable vector bundles are not lifted by instanton effects.  This has been demonstrated in the class of ``half-linear'' models via an analysis of the analytic structure of the spacetime superpotential in a paper by Beasley \& Witten \cite{Beasley:2003fx} and, in the more limited case of GLSMs, via a generalized Konishi anomaly argument by Basu \& Sethi\cite{Basu:2003bq}.  Due to the gauge anomaly and gauge variance of the classical Lagrangian, neither of these analyses directly apply to our torsion models; however, the basic structure of the Beasley-Witten argument obtains, which suggests that the vacuum is indeed stable to worldsheet instanton corrections.  

The basic ingredients in \cite{Beasley:2003fx} were that the spacetime superpotential is a holomorphic section of a simple line bundle; that poles can appear only if the instanton moduli space has a non-compact dimension along which worldsheet correlators can diverge; that a simple residue theorem ensures that the sum over all poles is zero; and that the worldsheet theory respect a linearly realized $(0,2)$ with non-anomalous $U(1)$ $R$-symmetry.  In the case of our TLSMs, the crucial step is verifying that the instanton moduli space is in fact compact; the rest appears to follow rather straightforwardly.

The instantons in our TLSM fall into two classes: those involving gauge fields coupled to torsion multiplets and those involving gauge fields coupled only to chiral multiplets.  The latter class is identical to those studied in \cite{Beasley:2003fx, Silverstein:1995re} and have compact moduli spaces for the same reasons; these correspond to the homologically non-trivial lifts of holomorphic curves on the base \CY.  The former is more subtle.  Recall that all that matters for the lifting of the massless vacuum are contributions to the chiral superpotential from BPS instantons.  Significantly, BPS instantons in the torsion sector must satisfy an unusual BPS equation
\[
\d\psi = \p_{+}\t + M^{a}v_{+a}=0.
\]
Since $v_{+a}$ is singular for an instanton background, instantons aligned along $M^{a}$ in $G$ do not have finite action, so we appear to have {\it no} instantons along the curve associated to $M^a$.  Actually, this makes a great deal of sense.  The one-form on $K3$ associated to $M^{a}v_{+a}$ is $\a_{M}$ (see appendix B); since $\a_{M}$ is not a globally-defined form, $\w_{M}=d\a_{M}$ -- the 2-form curvature of the $T^{2}$-bundle -- is non-trivial in $H^{2}(K3,\IZ)$.  However, the connection 1-form on $X$, $d\t+\pi^*\a_{M}$, \emph{is} a globally defined 1-form on $X$, so $d(d\t+\pi^*\a_{M})$ is trivial in $H^2(X,\IZ)$.  Thus, there is no 2-cycle in $X$ associated with this gauge field.  

Thus the BPS instantons of the TLSM are a refinement of the instantons of the base \CY, and the moduli space is consequently compact.  Elevating these heuristic arguments to a rigorous proof of the stability of the vacuum to instanton corrections does not appear impossible.  We leave a more thorough discussion of instantons in torsion sigma models, and a formal proof of the stability of the vacuum, to future work.

\section{Conclusions and Speculations}
\setcounter{equation}{0}

In this note, we have constructed gauged linear sigma models for \nK\ compactifications of the heterotic string with non-trivial background NS-NS 3-form $H$ satisfying the modified Bianchi identity, and we argued for the exact stability of their vacua to all orders and non-perturbatively in $\a'$.  This construction provides a microscopic definition of the \FY\ CFT and, via duality, for a related class of KST-like flux-vacua \cite{Kachru:2002he} involving non-trivial NS-NS and RR fluxes which stabilize various moduli in a fiducial \CY\ orientifold compactification.  

While motivated by the remarkable \FY\ construction, this construction is considerably more general, suggesting applications much richer than we have been able to cover explicitly.   For example, while we have focused on $K3$ bases for simplicity, it is completely straightforward to construct more general compactifications over higher-dimensional \CY\ bases, leading to 7 and 8 dimensional \nK\ compactifications corresponding to torsionful $G_{2}$ and $Spin(7)$ structure manifolds.  
It is also natural to try to apply the technology of the torsion multiplet to non-CY bases -- say, $dP_{8}$ -- by suitably adjusting the fibration structure.  
Perhaps the easiest cases to be studied are the type II examples in section 2; there is a rich story to be told there, including non-perturbative existence and a thorough study of the instanton structure of the theory.  We will return to all of these points in upcoming publications.

One area where our construction should be of particular use is in the study of the moduli spaces -- and hence low-energy phenomenology -- of \nK\ compactifications\cite{Becker:2005nb, Cyrier:2006pp}.  The necessary tools for analyzing the spectra of $(0,2)$ GLSMs have long been know \cite{Distler:1993mk} and can presumably be applied with minor modifications.  Relatedly, TLSMs should also provide a computationally effective tool to study the topological ring which was recently proved to exist for generic $(0,2)$-models \cite{Adams:2003zy,Adams:2005tc}, as well as the action of mirror symmetry on these stringy geometries.  In fact, the action of T-duality and mirror symmetry on these geometries is remarkably subtle -- for example, it is easy to check that the $T^{2}$ fibre on the \FY\ geometry is, in fact, self-dual, corresponding to a pair of $SU$(2) WZW models at level one.  What is the relation between the self-dual circles and the NS-NS flux?  Are these WZW models playing the anomaly-cancelling role of the WZW models in the $(0,2)$ Gepner model constructions of Berglund et al \cite{Berglund:1995dv}? Clearly, a great deal remains to be learned form these torsion linear sigma models, and from the CFTs to which they flow.

\vskip0.5in
\begin{center}
{\bf Acknowledgements}
\end{center}
\vskip0.1in
\noindent
We would like to thank
N.~Arkani-Hamed,
C.~Beasley, 
M.~Becker,
S.~Kachru,
A.~Lawrence,
J.~\mbox{McGreevy},
D.~Morrison,
N.~Seiberg,
A.~Simons,
A. Strominger,
W.~Taylor,
L.-S.~Tseng,
C.~Vafa,
S.-T.~Yau,
and
X.~Yin, 
warmly for helpful discussions.  A.A. thanks the Theory Group at the Weizmann Institute, the organizers and participants of the 2006 Amsterdam Workshop, and the Aspen Center for Physics for discussions and hospitality while this work was being completed.
The work of A.A. was supported by a Junior Fellowship from the Harvard Society of Fellows, and in part by the DOE under contract No. DE-FC02-94ER40818.  The work of M.E. and J.L. was supported by the DOE under contract No. DE-FG02-91ER40654.

\appendix

\renewcommand{\theequation}{\Alph{section}.\arabic{equation}}

\section{Review of $(0,2)$ and $(2,2)$ GLSMs}
\setcounter{equation}{0}

The following is a lightning review of the salient features of $(0,2)$ gauged linear sigma models; for more complete discussions see \cite{Witten:1993yc,Distler:1995mi}.  Our conventions and notation follow \cite{Hori:2003xxx}, with all factors of $\a'$ suppressed throughout the paper.
We take the \ZT\ superspace coordinates to be $(y^+, y^-,\t^+,\tb^+)$,
where $y^\pm = (y^0\pm y^1)$.  We begin with the gauge multiplet.

The right-moving gauge covariant superderivatives ${\cal{D}}_+,
{\overline{\cal{D}}}_+$, 
satisfy the algebra
\be
\CD_+^2 = \CDB_+^2 =0, \qquad  -\tfrac{i}{4}\{\ \!\CD_+,
\CDB_+ \}\ =\nabla_{+} =\p_+ +iQv_+ ,
\ee
where $Q$ is the charge of the field on which they act.  These imply that in a suitable basis we can identify
\be  
\CD_{+} =\frac{\partial}{\partial \theta^+} - 2i {\bar{\theta}}^+ \nabla_{+}, \qquad \CDB_{+} = -\frac{\partial}{\partial {\bar{\theta}}^+} + 2i\t^+  \nabla_{+},\qquad \CD_{-} = \p_{-} + \tfrac{i}{2} QV_{-}, \non
\ee
where $V_{\pm}$ are real vector superfields which transform under a gauge transformation with (uncharged) chiral gauge parameter $\CDB_{+}\LL=0$ as $\delta_{\LL} V_{-}$=$\p_-(\LL + \LB)$ and $\delta_{\LL} V_{+}$=$\frac{i}{2} (\LL-\LB)$; $\nabla_\pm$ are the usual gauge covariant derivatives.  This allows us to fix to Wess-Zumino gauge in which 
\be
V_{+} = \theta^+ {\bar{\theta}}^+ 2v_{+}  \qquad
V_{-} = 2v_{-} - 2i \theta^+ {\bar{\lambda}}_- -2i {\bar{\theta}}^+ \lambda_- +2 \theta^+ {\bar{\theta}}^+ D. \non
\ee
Note that $V_{-}$ contains a complex left-moving gaugino. Finally, the natural field strength is a fermionic chiral superfield, 
\be 
\U= 2[{\overline{\mathcal{D}}}_+ ,\CD_{-}] =\CDB_+(2\p_-V_{+} + i V_{-}) =
-2\{\lambda_- -i\tp(D+2iv_{+-}) - 2i\tp\bar{\tp}\p_+\lambda_- \}, 
\ee 
for which the natural action is
\be 
S_{\Upsilon} = - \frac{1}{8e^2}\int\!{\rm{d^2}}y\, d\t^+ d\bar{\t}^+ ~\overline{\U}\U =\frac{1}{e^2}
\int{\rm{d^2}}y\,
\left\{ 2v_{+-}^2  + 2i {\bar{\lambda}}_- \p_+ \lambda_-  +\frac{1}{2} D^2  \right\},
\ee 
where $d^2y = dy^0 dy^1$ and we use conventions where $\int d\t^+ \t^+ = \int \bar{\t}^+ d\bar{\t}^+ = 1$.

Matter multiplets are similarly straightforward.  A bosonic superfield satisfying $\CDB_{+}  \Phi =0$ is called a {\em chiral} supermultiplet and contains a complex scalar and a right-moving complex fermion $\Phi = \phi +\rt\theta^+\psi_+ - 2i\tp\tpb \nabla_{+}\phi$, and under gauge transformations $\Phi\to e^{-iQ(\LL+\LB)/2}\Phi$.  The gauge invariant Lagrangian is given by
\bea
 S_{\Phi} &=&  - i \int \,{\rm{d^2}}y\, {\rm{d^2}} \theta \
 {\overline{\Phi}} \CD_- \Phi, \\ &=&
 \int {\rm{d^2}}y\,  \Big\{ - \vert \nabla_{\alpha} \phi\vert^2 +
    2i{\bar{\psi}}_{+}\nabla_- \psi_{+} -i Q {\sqrt 2}
    {\bar{\phi}} \lambda_- \psi_{+} 
    +i Q {\sqrt 2} \phi
    {\bar{\psi}}_{+} {\bar{\lambda}}_-  + Q D \vert \phi \vert^2\Big\} , \non
\eea
where the metric is given by $\eta^{+-} = -2$.

Left-moving fermions transform in their own supermultiplet, the {\em fermi} supermultiplet, which satisfies the chiral constraint
\be\label{eq:almostchiral}
\overline{\cal{D}}_+ \G = \sqrt{2} E
\ee
and has component expansion $\G = \g_- -\rt\tp G
- 2i\tp \tpb \nabla_+ \g_{-} - \sqrt{2} \tpb E,$
where ${\overline{\cal{D}}}_+ E=0$ is a bosonic chiral superfield with the same gauge charge as $\Gamma$. 
The action for $\G$ is given by
 \bea \label{LF} S_{\G} &=& - \frac{1}{2}\int\!{\rm{d^2}}y\, d^2\t
 ~\overline{\G} \G \\
&=& \int {\rm{d^2}}y\, \left\{ 
    2i{\bar{\g}}_{-}  \nabla_+ \g_{-} + \vert G \vert^2 - \vert E
    \vert^2  - \left( \bar{\g}_- \frac{\partial E}{\partial
      \phi_i} \psi_{+i} + \bar{\psi}_{+i}
\frac{\partial \overline{E}}{\partial \bar{\phi}_i} \g_- \right) \right\}. \non
\eea

In general, we can add superpotential terms to our Lagrangian. Since these are integrals over a single supercoordinate, the superpotential can be written as a sum of fermi superfields $\G_m$ times holomorphic functions $J^m$ of the chiral superfields,
\bea \label{superpotential}
S_{\cal W} &=& \frac{1}{\sqrt{2}}\int\!{\rm{d^2}}y\,d\tp~ \G_m
J^m \vert_{{\bar\theta}^+=0} + {\rm
  h.c.},\\ &=& -\int{\rm{d^2}}y\, \left\{G_m J^m(\phi_i) + \g_{-m}
\psi_{+i}  \frac{\partial J^m}{
  \partial \phi_i} \right\}+ {\rm
  h.c.}.
\non
\eea
Since $\G_m$ is not an honest chiral superfield but satisfies
(\ref{eq:almostchiral}), we need to impose the condition
\be
E \cdot J=0
\ee
to ensure that the superpotential is chiral.
Finally, since $\U$ is a chiral fermion, we can also add an FI term of the form
\be 
\label{defineFI} 
S_{\rm FI} = \frac{it}{4}\int\!{\rm{d^2}}y\,
d\tp~ \U \vert_{{\bar{\theta}}^+ =0} + {\rm h.c.} = \int d^2y \left( -rD + 2\t v_{+-} \right)
\ee 
where $t=r+i\t$  is the complexified FI parameter.

\subsection{Our Canonical Example: $\V \to K3$}

Our canonical example begins with a vector bundle $\V \to S$ over a $K3$ hypersurface $S$ in a resolved weighted projective space $W\IP^{3}$.  The associated GLSM includes the gauge group $G=U(1)^{s}$ with $s$ gauge field-strengths $\U_{a}$, $(3+s)$ chiral scalars $\Phi_{i=1,...3+s}$ with charges $Q^{a}_{i}$, a set of $c$ neutral scalars $\Sigma_{A=1...c}$, a single chiral scalar $\Phi_{0}$ with charges $-d^{a}$, $r$ fermi multiplets $\Gamma_{m=1,...r}$ with charges $q^{a}_{m}$ satisfying the constraints $\overline{\cal{D}}_+\G_{m}=\rt \S_{A}E^{A}_{m}(\Phi)$, a single chiral fermion $\Gamma_{0}$ with charges $-m^{a}$, and spectators as needed to ensure vanishing of the one-loop tadpole for $D^{a}$ \cite{Distler:1993mk,Distler:1995mi}, all interacting according to the canonical Lagrangian density
\bea
{\cal L} & = & - \int d^{2}\t ~\left[    
     \frac{1}{8e_{a}^{2}}\overline{\U}_{a}\U_{a}  
+ \half\overline{\G}_{m}\G_{m}  
+ i\overline{\S}_{A}\p_{-} \S_{A}
+ i\overline{\Phi}_{i}(\p_{-}+ \tfrac{i}{2}Q^{a}_{i}V_{-a}) \Phi_{i}
\right] \non\\
&& + \frac{1}{\rt}\int d\t^{+} ~ [
      \G_{0} G(\Phi_{i})
+ \G_{m} \Phi_{0} J^{m}(\Phi_{i})
+ \frac{\rt}{4}it^{a}\U_{a}
]   +   {\rm h.c.}. \non
\eea
Integrating out the auxiliary fields results in a scalar potential
\[
U=\sum_{a}\frac{e_a^2}{2}
\left({\textstyle\sum_i} Q^{a}_{i}|\phi_{i}|^{2} -m^{a}|\phi_{0}|^{2}  -r^{a}\right)^{2}
+ |G(\phi)|^{2}  
+\sum_{m} \lp |\phi_{0}|^{2}|J^{m}(\phi)|^{2}
+ \left|{\textstyle\sum_A}\s_{A}E^{A}_{m}(\phi)\right|^{2} \rp.
\]
Non-singularity of the relevant geometric phase requires $G(\Phi_i)$ and $J^{m}(\Phi_i)$ to be transverse,
\bea
G=\frac{\p G}{\p\phi_1}=\frac{\p G}{\p\phi_2}=\cdots=0 \qquad &\Longleftrightarrow& \qquad \forall i:\phi_i=0,\non\\
G=J^1=J^2=\cdots=0\qquad &\Longleftrightarrow& \qquad \forall i:\phi_i=0.\non
\eea
In the relevant geometric phase of the K\"{a}hler cone, the Yukawa interactions
\bea
{\cal L}_{Yuk} &=&
-\bigg[\g_{-m}\lp \psi_{+0}J^{m} + \phi_{0}\psi_{+i}\frac{\p J^{m}}{\p\phi_{i}} \rp 
+ \sum_i \rt i Q_i^a\bar{\phi}_i\lambda_{-a}\psi_{+i}  \non\\
&& + \bar{\g}_{-m} \lp  \eta_{+A}  E^{A}_{m}  + \psi_{+i} \s_{A}\frac{\p E^{A}_{m}}{\p\phi_{i}} \rp 
+ \g_{-0}\psi_{+i}\frac{\p G}{\p \phi_i} \bigg]  +  {\rm h.c.} \non
\eea
give masses to various linear combinations of the right- and left-moving fermions.
The massless right-moving fermions couple to a bundle which fits into two exact sequences. For instance, for a single $U(1)$ we have
\bea\label{SESTS}
&&0 \to \mathcal{O}_{W\IP} \stackrel{Q_i\phi_i}{\longrightarrow} \oplus_i \mathcal{O}_{W\IP}(Q_i) \to T_{W\IP} \to 0 \label{eq:seq1}\\
&&0 \to T_S \to \left.T_{W\IP}\right|_S \stackrel{\p_{\phi_{i}}G}{\longrightarrow} \mathcal{O}_S(d) \to 0, \non
\eea
so the massless right-moving fermions couple to $T_{S}$.  
Similarly, the bundle $\V_{S}$ to which the massless left-moving fermions couple fits into a pair of short exact sequences,
\bea\label{SESVS}
&&0 \to \oplus_{A} \mathcal{O}_{W\IP} \stackrel{E^A_{m}}{\longrightarrow} \oplus_{m} \mathcal{O}(q_{m}) \to \V_{W\IP} \to 0 \label{eq:seq2}\\
&&0 \to \V_{S} \to \left.\V_{W\IP}\right|_S \stackrel{J^m}{\longrightarrow} \mathcal{O}(m) \to 0. \non
\eea

\subsection{GLSMs with $(2,2)$ Supersymmetry}

A special class of \ZT\ theories have enhanced $(2,2)$
supersymmetry. To describe these theories, we enlarge our superspace
by adding two fermionic coordinates,
$(y^+, y^-,\t^+,\tb^+, \t^-, \tb^-)$, and introduce
supercovariant derivatives
\be
\D_\pm = \pp{\t^\pm} - 2i\tb^\pm \p_\pm   \qquad   \DB_\pm = -\pp{\tb^\pm} + 2i\t^\pm\p_\pm.
\ee
Unlike the \ZT\ case, there are two kinds of \TT\ chiral
multiplets, {\em chiral} multiplets satisfying
\be
\DB_+ \Phi_{2,2} = \DB_- \Phi_{2,2} =0,
\ee
and {\em twisted chiral} multiplets satisfying
\be
\DB_+ Y_{2,2} = \D_- Y_{2,2} =0.
\ee
Both have the field content of one \ZT\ chiral and one \ZT\ fermi multiplet,
\be
\Phi_{2,2} = \Phi +\rt\tm\G_{-}  - 2i\tm\tmb \p_{-}\Phi  \qquad
Y_{2,2} = Y +\rt\tb^{-}F  + 2i\tm\tmb \p_{-} Y. \non
\ee
The $(2,2)$ vector superfield, $V_{2,2}$, whose field strength
is a twisted chiral multiplet $\S=\frac{1}{\rt}\overline{D}_+ D_- V_{2,2}$, is built out of an uncharged \ZT\ chiral multiplet $\S_0$ and a \ZT\ vector
multiplet $V_{\pm}$ as
\be
V_{2,2} = V_+ + \tm\bar{\t}^- V_- + \rt\bar{\t}^+\tm\S_0 + \rt\bar{\t}^-\tp\overline{\S}_0 ,
\ee
where $\S_0 = \sigma - i\rt\tp\bar{\lambda}_+ - 2i\tp\bar{\tp}\p_+\sigma$ for agreement with \cite{Witten:1993yc}, and $\delta_g V_{2,2} = \frac{i}{2}\left(\LL_{2,2} - \LB_{2,2}\right)$.  The standard FI-term is
\[
{\cal L}_{FI} = \frac{-t}{2\rt} \int d\tp d\tmb \S + \hc = -rD + 2\t v_{+-},
\]
where $t=r+i\t$.

Lastly, we note that a $(2,2)$ chiral multiplet with $U(1)$ charge $Q$
reduces to a charged \ZT\ chiral multiplet $\Phi$ and a charged
fermi multiplet $\G$ satisfying
\[
\CDB_+ \G = \sqrt{2} E
\]
in $(0,2)$ notation, and where $E$ is given by 
\be
E = {\sqrt 2} Q \Sigma_0 \Phi.
\ee
We will omit the subscripts ``2,2'' in the main text, as it should always be clear from the context to which \susy\ we refer.

\section{The Fu-Yau Geometry}\label{GPFY}
\setcounter{equation}{0}

\subsection{Supersymmetry Constraints}

Consider compactification of the heterotic string on a 6-dimensional manifold, $X$.  Preserving $\N$=1 \susy\ in 4d requires that $X$ admit a nowhere vanishing spinor, $\eta$.  This immediately implies that $X$ admits an almost complex structure.  The existence of a nowhere-vanishing spinor on an almost-complex 3-fold implies that the frame bundle admits a connection of $SU(3)$ holonomy -- \ie\ that $X$ is a special-holonomy manifold with $SU(3)$-structure.  However, the connection of special holonomy need not be the Levi-Civita connection, and in general the nowhere-vanishing spinor is {\it not} annihilated by the metric connection, $\nabla_{g}$, but by a (unique) torsionful connection, 
\[
(\nabla_{g} + H) \eta = 0.
\]
$H$ is called the intrinsic torsion of the $SU(3)$-structure.  In the special case $H=0$, when the nowhere-vanishing spinor is covariantly constant according to the metric connection, $X$ admits a metric of $SU(3)$ holonomy and is thus Calabi-Yau.

$\N=1$ supersymmetry in 4d further requires the vanishing of the supersymmetry variations of the gravitino, dilatino, and gaugino.  Together with the Jacobi identity for the resulting superalgebra, these constraints imply that $X$ admits an {\it integrable} complex structure, a nowhere-vanishing Hermitian metric corresponding to a globally-defined Hermitian (1,1)-form, $J_{a\bar{b}}=\eta^{\dagger}\Gamma_{a\bar{b}}\eta$, a nowhere-vanishing holomorphic (3,0)-form, $\Omega_{abc}=\eta^{\dagger}\Gamma_{abc}\eta$, and comes equipped with a Hermitian-Yang-Mills gauge field, $F^{(2,0)}=F^{(0,2)}=F_{mn}J^{mn}=0.$  They also imply that 
\[
H = i(\pb-\p)J,
\]
so $H$ is the obstruction to $X$ being K\"{a}hler.  Instead, $X$ is {\it conformally balanced}, 
\[
d(e^{-2\phi}J\wedge J)=0,
\]
where $\phi$ is the Einstein-frame dilaton.  While more complicated than the simple $H=0$ Calabi-Yau case, the general $X$ would so far appear to be on a similar footing.

The Green-Schwarz anomaly completely changes the story.  Including the one-loop gravitational correction, the vanishing of the anomaly implies
\[
dH = \a' \lp \mathrm{tr} R\wedge R - \tr F\wedge F \rp ,
\]
%
where $R$ is the curvature of the Hermitian connection on $X$.  This changes the story in several dramatic ways.  First, since the left and right hand sides of this equation scale inhomogenously in the global conformal mode of the metric, any solution to this equation has fixed volume modulus.  Crucially, this means any solution to this equation does {\it not} have a large radius limit, so supergravity perturbation theory has a finite, fixed expansion parameter and must be taken with a sizeable grain of salt.  Secondly, this equation is spectacularly nonlinear, so even proving the existence of solutions is a profoundly difficult problem.

Happily, in at least one special case there exists an existence proof by Fu and Yau for solutions to the full set of conditions outlined above, including the anomaly equation, analogous to Yau's proof of the existence of a Ricci-flat K\"{a}hler metric on manifolds of $SU(3)$-holonomy.  Unlike the Yau proof of the Calabi conjecture, however, the Fu-Yau proof begins with a very specific Ansatz for the metric, torsion, and holomorphic 3-form \cite{Fu:2006vj}.

\subsection{GP Manifolds and the FY Compactification}

The underlying manifold satisfying all of the supersymmetry constraints unrelated to the gauge bundle was first constructed by Goldstein and Prokushkin \cite{Goldstein:2002pg}.  Their solution involved constructing the complex 3-fold as a $T^2$ bundle over a $T^4$ or $K3$ base.  Fu and Yau \cite{Fu:2006vj} used this underlying manifold and constructed a gauge bundle satisfying the remaining supersymmetry constraints as well as the modified Bianchi identity, which was a monumental accomplishment since it is a complicated differential equation.  We start by explaining the GP manifold.


Let $S$ be a complex Hermitian 2-fold and choose\footnote{Actually, Goldstein and Prokushkin only required that $\omega_P + i\omega_Q$ have no $(0,2)$-component, but Fu and Yau used the restriction that we have stated.}
\be
\frac{\omega_P}{2\pi},\frac{\omega_Q}{2\pi}\in H^2(S;\bb{Z}) \cap \Lambda^{1,1}T^*_S .
\ee
where $\omega_P$ and $\omega_Q$ are anti self-dual. Being elements of integer cohomology, there are two $\mathbb{C}^*$-bundles over $S$, call them $P$ and $Q$, whose curvature 2-forms are $\omega_P$ and $\omega_Q$, respectively.  We can then restrict to unit-circle bundles $S_P^1$ and $S_Q^1$ of $P$ and $Q$ respectively, and take the product of the two circles over each point in $S$ to form a $T^2$ bundle over $S$ which we will refer to as $X$ ($T^2\rightarrow X \stackrel{\pi}{\rightarrow}  S$).

Given this setup, Goldstein and Prokushkin showed that if $S$ admits a non-vanishing, holomorphic $(2,0)$-form, then $X$ admits a non-vanishing, holomorphic $(3,0)$-form.  Furthermore, they showed that if $\omega_P$ or $\omega_Q$ are nontrivial in cohomology on $S$, then $X$ admits \emph{no} K\"ahler metric.  They constructed the non-vanishing holomorphic $(3,0)$-form and a Hermitian metric on $X$ from data on $S$.

The curvature 2-form $\omega_P$ determines a non-unique connection $\nabla$ on $S^1_P$ (and similarly for $\omega_Q$ on $S^1_Q$).  A connection determines a split of $T_X$ into a vertical and horizontal subbundle -- the horizontal subbundle is composed of the elements of $T_X$ that are annihilated by the connection $1$-form, the vertical subbundle is then, roughly speaking, the elements of $T_X$ tangent to the fibres.  Over an open subset $U\subset S$, we have a local trivialization of $X$ and we can use unit-norm sections, $\xi\in\Gamma(U;S^1_P)$ and $\zeta\in\Gamma(U;S^1_Q)$, to define local coordinates for $z\in U\times T^2$ by
\be
\label{eqn:coords}
z = (p,e^{i\theta_P}\xi(p),e^{i\theta_Q}\zeta(p)) , 
\ee
where $p = \pi(z)\in U$.  The sections $\xi$ and $\zeta$ also define connection 1-forms via
\be
\nabla \xi = i\alpha_P\otimes \xi      \qquad     \textrm{and}     \qquad     \nabla \zeta = i\alpha_Q\otimes\zeta,
\ee
where $\omega_P = d\alpha_P$ and $\omega_Q = d\alpha_Q$ on $U$, and the $\alpha_i$ are necessarily real to preserve the unit-norms of $\xi$ and $\zeta$.

The complex structure is given on the fibres by $\partial_{\theta_P}\rightarrow\partial_{\theta_Q}$ and $\partial_{\theta_Q}\rightarrow -\partial_{\theta_P}$ while on the horizontal distribution it is induced by projection onto $S$.\footnote{Actually, this just gives an almost complex structure, but Goldstein and Prokushkin proved that it is integrable \cite{Goldstein:2002pg}}  Given a Hermitian $2$-form $\omega_S$ on $S$, the 2-form
\be
\label{eqn:hermitian form}
\omega_u = \pi^*\left(e^{u} \omega_M\right) + (d\theta_P+\pi^*\alpha_P)\wedge(d\theta_Q+\pi^*\alpha_Q),
\ee
where $u$ is some smooth function on $S$, is a Hermitian $2$-form on $X$ with respect to this complex structure.  The connection $1$-form
\be
\vartheta = (d\theta_P+\pi^*\alpha_P) + i(d\theta_Q+\pi^*\alpha_Q)
\ee
annihilates elements of the horizontal distribution of $T_X$ while reducing to $d\theta_P+id\theta_Q$ along the fibres.  These data define the complex Hermitian $3$-fold $(X,\omega_u)$, which we call the GP manifold \cite{Goldstein:2002pg}.  Explicitly,
\bea
ds^{2}_{X} &=& \pi^*\left(e^{u}ds^{2}_{S}\right) + (d\theta_P+\pi^*\a_P)^{2} + (d\theta_Q+\pi^*\a_Q)^{2} \non \\
J_{X} &=& \pi^*\left(e^{u}J_{S}\right) + \half \vartheta\wedge\bar{\vartheta}  \non \\
\Omega_{X} &=& \pi^*\left(\Omega_{S}\right)\wedge\vartheta  \non \\
H &=& \sum_{i=P,Q} (d\theta_{i}+\pi^*\a_{i})\wedge\pi^*\w_{i},  \non 
\eea
where $\Omega_{S}$ is the nowhere-vanishing, holomorphic $(2,0)$-form on $S$ ($K3$ or $T^4$).  It is straightforward check that all the supersymmetry constraints are satisfied by this Ansatz, however for a valid heterotic compactifications a gauge bundle still needed to be constructed to satisfy the Bianchi identity.


Fu and Yau undertook the more difficult problem of proving the existence of gauge bundles over the GP manifold with Hermitian-Yang-Mills connections satisfying the Bianchi identity (\ref{eq:HBI}).  They took the Hermitian form (\ref{eqn:hermitian form}) and converted the Bianchi identity into a differential equation for the function $u$.  Under the assumption
\be
\label{eqn:u assumptions}
\left(\int_{K3} e^{-4u}\frac{\omega_{K3}^2}{2} \right)^{1/4} \ll 1 = \int_{K3} \frac{\omega_{K3}^2}{2} ,
\ee
they showed that there exists a solution $u$ to the Bianchi identity for \emph{any} compatible choice of gauge bundle $\V_{X}$ and curvatures $\omega_P$ and $\omega_Q$ such that the gauge bundle $\V_{X}$ over $X$ is the pullback of a stable, degree 0 bundle $\V_{K3}$ over $K3$, $\V_{X} = \pi^*\V_{K3}$ \cite{Fu:2006vj}; this is what we call the \FY\ geometry.  

Note that by a ``compatible'' choice of gauge bundle and $\omega_i$'s we mean the following: choose the gauge bundle $\V_{X}$ and the curvature forms to satisfy the integrated Bianchi identity
\be\label{eq:topcond}
\chi(S) - \tr F^2 = \int_S \sum_i \omega_i^2 .
\ee
In particular, note that the right-hand side and $\tr F^2$ are manifestly non-negative, since $*_S F=-F$ and $F$ is anti-Hermitian. Hence, the only possible solution for a $T^4$ base is to take the gauge bundle \emph{and} the $T^2$ bundle to be trivial, leaving us with a Calabi-Yau solution $T^2\times T^4$ \cite{Becker:2006et,Fu:2006vj}.  This is in agreement with arguments from string duality ruling out the Iwasawa manifold as a solution to the heterotic supersymmetry constraints \cite{Gauntlett:2003cy}.

\bibliography{TLSMrefs}
\bibliographystyle{hunsrt}

\end{document}